\documentclass[aps,prx,reprint,superscriptaddress,longbibliography,floatfix,showkeys]{revtex4-2}
\usepackage{amsmath}
\usepackage{amssymb} 
\usepackage{fix-cm}
\usepackage[defaultsups]{newtxtext}
\usepackage[varg]{newtxmath}
\usepackage{graphicx}
\usepackage{xurl} 
\usepackage[
  colorlinks=true,
  linkcolor=blue,
  citecolor=blue,
  urlcolor=blue
]{hyperref}
\usepackage{booktabs}
\renewcommand{\arraystretch}{1.15}
\usepackage{placeins}

\usepackage{tikz}
\usetikzlibrary{calc,positioning,fit,arrows.meta}

\usepackage{siunitx}
\newcommand{\tblcell}[2]{\parbox[t]{#1}{\raggedright\strut #2\strut}}

\newcommand{\safeincludegraphics}[2][]{%
  \IfFileExists{#2}{%
    \includegraphics[#1]{#2}%
  }{%
    \fbox{\parbox[c][0.18\textheight][c]{0.95\linewidth}{\centering Missing figure file\\\texttt{\detokenize{#2}}}}%
  }%
}
\newcommand{\inlinegraphicorfallback}[2][]{%
  \IfFileExists{#2}{\includegraphics[#1]{#2}}{\textsf{[graphic]}}%
}

\begin{document}

\title{Metrology for Quantum Hardware Standardization\\
\textemdash~Charting a Pathway: A Strategic Review~\textemdash}

\author{Nobu-Hisa Kaneko}
\email{nobuhisa.kaneko@aist.go.jp}
\affiliation{AIST, Global Research and Development Center for Business by Quantum-AI Technology (G-QuAT) and National Metrology Institute of Japan (NMIJ), 1-1-1 Umezono, Tsukuba, Ibaraki 305-8568, Japan}

\date{Received February 26, 2026}
\begin{abstract}
Advances in quantum mechanics have long underpinned metrology by enabling practical realizations of units through quantum effects. With the 2019 SI revision, traceability is anchored in defined fundamental constants, reinforcing the quantum-mechanical basis of modern standards. In parallel, quantum technologies are transitioning from laboratory science to engineering and early industrial deployment, bringing familiar pressures for integration, reliability, cost reduction, supply-chain formation, and standardization. The direction of benefit is thus reversing: metrology and precision measurement are becoming enabling infrastructure for the industrialization of quantum technologies.

Against this backdrop, this paper surveys the metrology and precision-measurement capabilities required across representative quantum-computing modalities and identifies where electrical and related metrology can contribute to the development, characterization, and reliable operation of quantum hardware. We then discuss cross-cutting measurement needs and standardization opportunities that recur across platforms, and note how similar frameworks can extend to emerging quantum-sensing applications.

\end{abstract}

\maketitle

\section{Introduction}
\label{sec:intro}

The discovery of the Josephson effect in 1962~\cite{Josephson1962} and the quantum Hall effect in 1980~\cite{vonKlitzing1980}, together with the subsequent development, maintenance, and internationally equivalent realization of quantum voltage and quantum resistance standards based on these effects, represent major milestones in which physics---especially quantum mechanics---has made a profound and direct contribution to the real economy. Similar developments have occurred in many other areas, including time and frequency standards, where laser technology, optical frequency combs, and atomic trapping have been central, and mass standards, where the electrical determination of the Planck constant has played a key role.

At the same time, today's quantum computers also arise directly from the fundamental principles of quantum mechanics and have recently attracted tremendous attention. In particular, the large-scale, sustained, and strategically coordinated public investments by advanced economies---and the emergence of numerous high-calibre start-ups that have attracted substantial private capital---are remarkable.

Research that, some ten years ago, focused largely on small, fundamental quantum systems and their potential use in computation, and was regarded as a topic for the relatively distant future, has in the past five years rapidly become a focus of industrial interest.

Several roadmaps now anticipate the realization of small-scale fault-tolerant quantum computers across most quantum-computing modalities within the coming five years, suggesting that engineering-driven development will accelerate further.

\begin{figure}[htbp]
      \centering %
      \safeincludegraphics[width=\columnwidth]{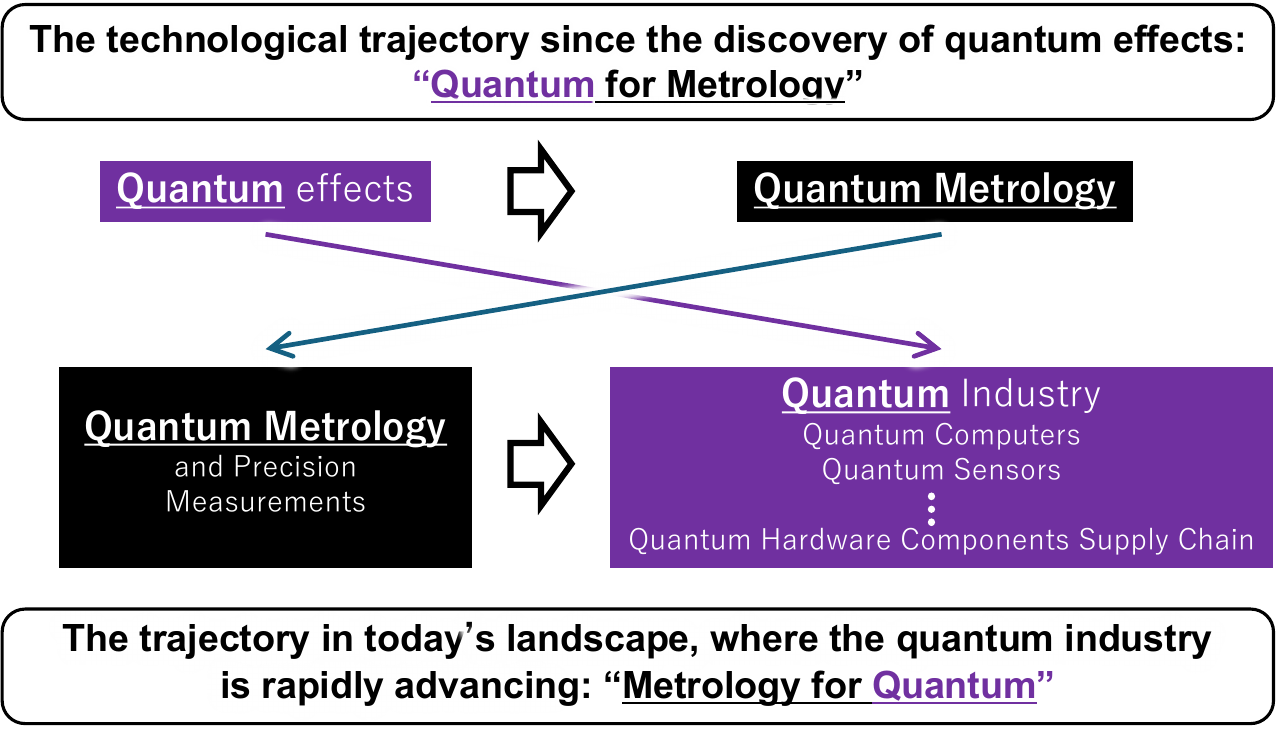}
      \caption{Until the 2019 SI revision, quantum metrological standards were established by leveraging quantum effects. Today, those quantum metrological standards and the precision measurement technologies built on them are poised to contribute to the quantum industry.}
      \label{fig:Standards_to_Quantum}
\end{figure}

In this context, metrology---long propelled by the exploitation of quantum effects (i.e., ``Quantum for Metrology'')---is now poised to make a major contribution to the development of quantum computers and sensors (i.e., ``Metrology for quantum'')~\cite{WS-NMI-Q}. This relationship is illustrated schematically in Fig.~\ref{fig:Standards_to_Quantum}.

This paper is organized as follows. Section~\ref{sec:NMI-Q_JTC3} outlines the pre-standardization and standardization landscape, focusing on NMI-Q and IEC/ISO JTC~3. Section~\ref{sec:QCModalities} then briefly surveys representative quantum-computing modalities to motivate both cross-cutting and modality-specific measurement challenges. Section~\ref{sec:crosscutting} discusses cross-cutting measurement needs and standardization opportunities for hardware components and subsystems, with brief remarks on how the same framing can inform emerging quantum-sensing applications. Finally, Section~\ref{sec:summary} summarizes the emerging convergence toward shared, modality-agnostic component-development platforms alongside the continuing need for modality-specific enabling technologies.

\section{Pre-standardization and Standardization Pathways for Quantum Technologies: NMI-Q and IEC/ISO JTC 3}\label{sec:NMI-Q_JTC3}
As quantum technologies scale from laboratory prototypes to engineered systems, comparable characterization and a shared measurement language become essential for reliability, supply-chain formation, and system integration. This section provides the standardization framing used throughout the paper by describing how pre-standardization efforts (e.g., within NMI-Q) interface with formal international standardization through organizations such as the IEC (International Electrotechnical Commission) and ISO (International Organization for Standardization). It also clarifies the scope and terminology needed to connect modality-specific requirements to the cross-cutting metrics, test methods, and traceability pathways discussed in later sections.

In traditional small-scale quantum systems, it was taken for granted that qubit devices and the components of their surrounding environment would be fabricated and selected by hand, with researchers devoting substantial time and relying heavily on their own skills and experience. As system sizes grow, however, such person-dependent approaches are no longer sustainable. Going forward, these devices and component families must be characterized using systematic methods with high reliability and reproducibility and, where appropriate, managed under quality-control processes aligned with relevant international standards. The evaluation and selection processes themselves must also transition to more industrialized procedures suitable for large-scale deployment. This kind of evolution is commonplace in other industrial sectors, and it is an essential step as quantum computers scale up and move toward genuine industrial utilization. Along this path, integration, system development, supply-chain formation, cost reduction, and international standardization will all be indispensable.

\begin{table}[t]
\centering
\caption{Organizational structure of IEC/ISO JTC 3. It comprises eight Working Groups, two Project Teams, two Advisory Groups, and two Ad Hoc Groups. In addition, liaisons have been established with a broad range of application-oriented technical committees (TCs), see Table~\ref{tab:Liaisons_2026Feb} (as of June 2026).}
\label{tab:JTC3_structure_2026Feb}
{\scriptsize
\setlength{\tabcolsep}{2pt}
\renewcommand{\arraystretch}{1.05}
\begin{tabular}{@{}lll@{}}
\toprule
\tblcell{0.75cm}{\textbf{Group}} & \tblcell{4.95cm}{\textbf{Scope}} & \tblcell{1.95cm}{\textbf{Convener}} \\
\midrule
\multicolumn{3}{@{}l}{\textbf{Working Groups}}\\
\tblcell{0.75cm}{WG 9} & \tblcell{4.95cm}{Terminology and quantities} & \tblcell{1.95cm}{John Devaney (GB)} \\
\tblcell{0.75cm}{WG 10} & \tblcell{4.95cm}{Quantum sensors} & \tblcell{1.95cm}{Joon-Shik Park (KR)} \\
\tblcell{0.75cm}{WG 11} & \tblcell{4.95cm}{Quantum computing supply chain} & \tblcell{1.95cm}{Austin Lin (US)} \\
\tblcell{0.75cm}{WG 12} & \tblcell{4.95cm}{Quantum computing benchmarking} & \tblcell{1.95cm}{Masahiro Horibe (JP)} \\
\tblcell{0.75cm}{WG 13} & \tblcell{4.95cm}{Quantum random number generators} & \tblcell{1.95cm}{Minghan Li (CN)} \\
\tblcell{0.75cm}{WG 14} & \tblcell{4.95cm}{Quantum enabling technology} & \tblcell{1.95cm}{David Balslev-Harder (DK)} \\
\tblcell{0.75cm}{WG 15} & \tblcell{4.95cm}{Quantum computing terminology and quantities} & \tblcell{1.95cm}{Razieh Annabestani (CA)} \\
\tblcell{0.75cm}{WG 16} & \tblcell{4.95cm}{Quantum communication} & \tblcell{1.95cm}{Vicente Mart\'{i}n (ES)} \\
\addlinespace[0.2em]
\multicolumn{3}{@{}l}{\textbf{Project Teams}}\\
\tblcell{0.75cm}{PT 18157} & \tblcell{4.95cm}{PT ISO/IEC TR 18157} & \tblcell{1.95cm}{Jingjing Wang (CN)} \\
\tblcell{0.75cm}{PT 63622} & \tblcell{4.95cm}{Quantum photonics vocabulary} & \tblcell{1.95cm}{Joshua Bienfang (US)} \\
\addlinespace[0.2em]
\multicolumn{3}{@{}l}{\textbf{Advisory Groups}}\\
\tblcell{0.75cm}{AG 1} & \tblcell{4.95cm}{Strategic planning} & \tblcell{1.95cm}{Michael Egan (AU)} \\
\tblcell{0.75cm}{AG 8} & \tblcell{4.95cm}{Chair's advisory group} & \tblcell{1.95cm}{Haeseong Lee (KR)} \\
\addlinespace[0.2em]
\multicolumn{3}{@{}l}{\textbf{Ad-Hoc Groups}}\\
\tblcell{0.75cm}{ahG 4} & \tblcell{4.95cm}{Quantum communication} & \tblcell{1.95cm}{Seong Su Park (KR)} \\
\tblcell{0.75cm}{ahG 5} & \tblcell{4.95cm}{Quantum computing and simulation} & \tblcell{1.95cm}{Michael Egan (AU)} \\
\bottomrule
\end{tabular}
}
\vspace{2mm}

\begin{minipage}{0.9\linewidth}
\footnotesize
Note: Joint Subcommittee JSC 3A “Quantum Computing” was established following ratification by the IEC SMB and ISO TMB at the end of April 2026, and the secretariat of the subcommittee was allocated to AFNOR, France.
\end{minipage}

\end{table}

\begin{table}[htbp]
\centering
\caption{IEC/ISO JTC 3 liaisons with other IEC TCs, ISO TCs, and other SDOs (as of June 2026).}
\label{tab:Liaisons_2026Feb}
{\scriptsize
\setlength{\tabcolsep}{2pt}
\renewcommand{\arraystretch}{1.03}
\begin{tabular}{@{}ll@{}}
\toprule
\tblcell{1.55cm}{\textbf{Group}} & \tblcell{5.85cm}{\textbf{Scope}} \\
\midrule
\multicolumn{2}{@{}l}{\textbf{Internal IEC Liaisons}}\\
\tblcell{1.55cm}{TC 46} & \tblcell{5.85cm}{Cables, wires, waveguides, RF connectors, RF and microwave passive components and accessories} \\
\tblcell{1.55cm}{TC 47} & \tblcell{5.85cm}{Semiconductor devices} \\
\tblcell{1.55cm}{TC 62} & \tblcell{5.85cm}{Medical equipment, software, and systems} \\
\tblcell{1.55cm}{TC 76} & \tblcell{5.85cm}{Optical radiation safety and laser equipment} \\
\tblcell{1.55cm}{TC 86} & \tblcell{5.85cm}{Fibre optics} \\
\tblcell{1.55cm}{TC 90} & \tblcell{5.85cm}{Superconductivity} \\
\tblcell{1.55cm}{TC 108} & \tblcell{5.85cm}{Safety of electronic equipment within the field of audio/video, information technology and communication technology} \\
\tblcell{1.55cm}{TC 113} & \tblcell{5.85cm}{Nanotechnology for electrotechnical products and systems} \\
\tblcell{1.55cm}{ISO/IEC JTC 1} & \tblcell{5.85cm}{Information technology} \\
\tblcell{1.55cm}{ISO/IEC JTC 1/SC 6} & \tblcell{5.85cm}{Telecommunications and information exchange between systems} \\
\tblcell{1.55cm}{ISO/IEC JTC 1/SC 7} & \tblcell{5.85cm}{Software and systems engineering} \\
\tblcell{1.55cm}{ISO/IEC JTC 1/SC 22} & \tblcell{5.85cm}{Programming languages, their environments and system software interfaces} \\
\tblcell{1.55cm}{ISO/IEC JTC 1/SC 27} & \tblcell{5.85cm}{Information security, cybersecurity and privacy protection} \\
\addlinespace[0.2em]
\tblcell{1.55cm}{ISO/IEC JTC 1/SC 42} & \tblcell{5.85cm}{Artificial Intelligence} \\
\addlinespace[0.2em]
\multicolumn{2}{@{}l}{\textbf{Liaisons with ISO TCs}}\\
\tblcell{1.55cm}{ISO/TC 172} & \tblcell{5.85cm}{Optics and photonics} \\
\tblcell{1.55cm}{ISO/TC 201} & \tblcell{5.85cm}{Surface chemical analysis} \\
\tblcell{1.55cm}{ISO/TC 206} & \tblcell{5.85cm}{Fine ceramics} \\
\tblcell{1.55cm}{ISO/TC 220} & \tblcell{5.85cm}{Cryogenic vessels} \\
\tblcell{1.55cm}{ISO/TC 229} & \tblcell{5.85cm}{Nanotechnologies} \\
\addlinespace[0.2em]
\multicolumn{2}{@{}l}{\textbf{Liaisons with Other SDOs (Liaison A)}}\\
\tblcell{1.55cm}{EC} & \tblcell{5.85cm}{European Commission} \\
\tblcell{1.55cm}{ETSI} & \tblcell{5.85cm}{European Telecommunications Standards Institute} \\
\tblcell{1.55cm}{IAF} & \tblcell{5.85cm}{International Accreditation Forum} \\
\tblcell{1.55cm}{IEEE} & \tblcell{5.85cm}{Institute of Electrical and Electronics Engineers} \\
\tblcell{1.55cm}{ITU-T} & \tblcell{5.85cm}{International Telecommunication Union - Telecommunication Standardization Bureau} \\
\bottomrule
\end{tabular}
}
\end{table}

In this context, the international standardization of quantum technologies being carried out by IEC/ISO Joint Technical Committee~3 (JTC~3) on Quantum Technologies~\cite{JTC3}, jointly established by the IEC and ISO, represents an extremely important development. Within JTC~3, in addition to Advisory Group~1 (Strategic Planning), several Ad Hoc groups (ahGs) have been actively engaged in discussions since 2025. Building on the outcomes of these ahGs, formal Working Groups (WGs) have subsequently been established to address necessary and relevant topics. As of February~2026, IEC/ISO JTC~3 comprises eight Working Groups---Terminology and quantities; Quantum sensors; Quantum computing supply chain; Quantum computing benchmarking; Quantum random number generators; Quantum enabling technology; Quantum computing terminology and quantities; and Quantum communication---along with two Project Teams (PT ISO/IEC TR~18157 and Quantum photonics vocabulary), two Advisory Groups (Strategic Planning and the Chair's Advisory Group), and two ahGs (Quantum communication and Quantum computing and simulation). The current organizational structure of IEC/ISO JTC~3 is shown in Table~\ref{tab:JTC3_structure_2026Feb}. 

Reflecting the cross-cutting nature of quantum technologies, JTC~3 has also established liaisons with a broad range of application-oriented technical committees (TCs)---thirteen IEC internal TCs and five ISO TCs---as well as five other standards development organizations (SDOs). For details, see Table~\ref{tab:Liaisons_2026Feb}. It is evident that a broad range of standardization activities is underway and that liaison relationships have been established accordingly. In addition to these formal liaisons, information exchange is also taking place at the working level, and the number of such interactions is expected to increase further.

These groups examine, within their respective domains, what international standards and technical reports are required, and address issues such as appropriate performance metrics, test methods, and harmonized terminology. Furthermore, as seen in JTC~3 and in the \emph{Standardization Roadmap on Quantum Technologies} produced by European CEN/CENELEC JTC~22 ``Quantum Technologies''~\cite{CENCENELEC}, pre-standardization activities---such as the development of glossaries, reference architectures, benchmark problem sets, and round-robin testing---are becoming active in various regions. It is fair to say that we have entered a crucial phase of foundational work that will underpin future international quantum-technology standards.

NMI-Q~\cite{NMI-Q} or~\raisebox{-0.1\height}{\inlinegraphicorfallback[height=1.2em]{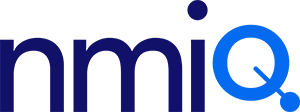}} is a new international initiative for pre-standardization that brings together the national metrology institutes (NMIs) of the G7 countries and Australia to accelerate the safe and effective development of quantum technologies. The initiative was politically endorsed in June~2025 in Kananaskis, Alberta, where G7 leaders adopted the ``Kananaskis Common Vision for the Future of Quantum Technologies'' and explicitly committed to ``intensify collaboration between trusted national measurement institutes, including via the NMI-Q initiative'' to advance essential measurement and testing work~\cite{KananaskisG7}. Beyond the founding members, other NMIs, academic institutions, and industrial or research organizations that share the same objectives are strongly encouraged to participate in NMI-Q activities, subject to the consensus of the full Steering Committee.

\begin{figure}[htbp]
      \centering %
      \safeincludegraphics[width=\columnwidth]{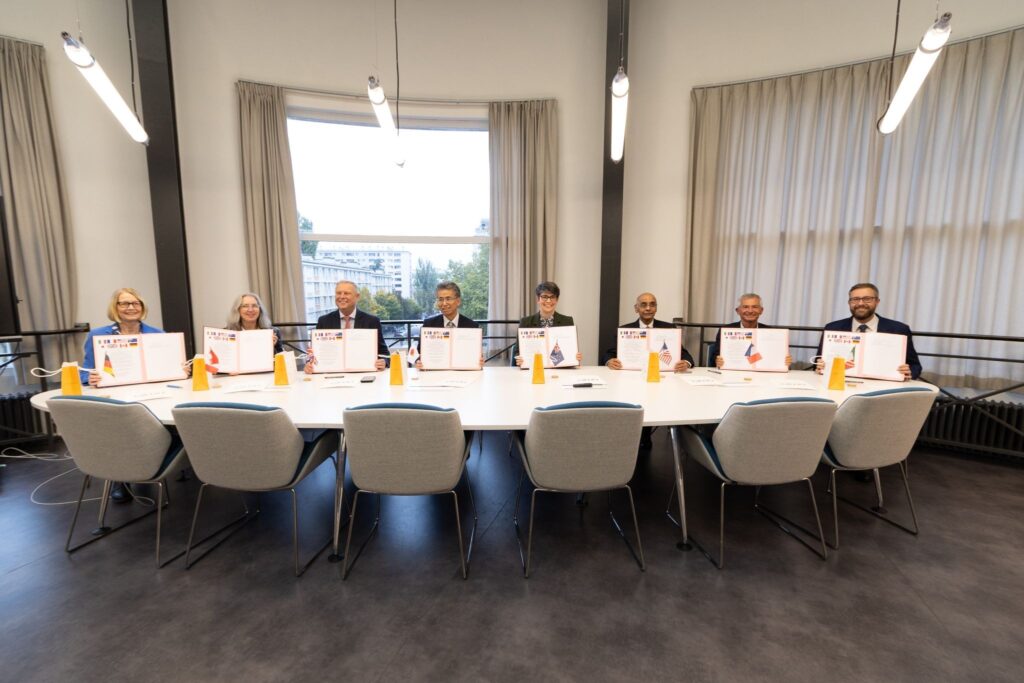}
      \caption{NMI-Q MoU signature event at Laboratoire national de m\'{e}trologie et d'essais (LNE), Paris on October 15, 2025. Photo credit: LNE.}
      \label{fig:NMI-Q_SIGNATURE}
\end{figure}

Building on this mandate NMI-Q was formally established through a Memorandum of Understanding (MoU) signed in Paris on 15~October~2025 at the premises of Laboratoire national de m\'{e}trologie et d'essais (LNE), the French national metrology institute. A photograph of the signature event is shown in Fiigure~\ref{fig:NMI-Q_SIGNATURE}. The founding members are LNE (France); the National Measurement Institute, Australia (NMIA); the National Research Council Canada (NRC); the Physikalisch-Technische Bundesanstalt (PTB), Germany; the Istituto Nazionale di Ricerca Metrologica (INRiM), Italy; the National Metrology Institute of Japan (NMIJ); the National Physical Laboratory (NPL), United Kingdom; and the National Institute of Standards and Technology (NIST), United States.

These eight NMIs form the NMI-Q Steering Committee, which sets the initiative's strategic direction. The public launch followed on 3~November~2025 at the ``Quantum Metrology: from foundations to the future'' event hosted by NPL in Teddington, UK, during the International Year of Quantum.

NMI-Q's mission is to support the development and adoption of quantum technologies---spanning computing, sensing, and communication---by creating harmonized measurement methods, sharing best practices, and contributing to international standards. The initiative recognizes that a globally functioning quantum market requires reliable, interoperable, and scalable measurement procedures. Without agreed performance metrics, traceability chains, and validation protocols, it is difficult to compare devices, certify products, or build trust among users and investors.

Operationally, NMI-Q is organized around Technical Working Areas (TWAs), which are the core collaborative themes of the program. Each TWA brings together experts from the founding NMIs and approved partners to address key challenges in quantum measurement science. Under each TWA, specific technical projects are established to deliver focused outcomes, ranging from pre-standardization research and interlaboratory comparisons (e.g., round-robin tests) to the development of best-measurement-practice guides and input documents for formal standards bodies or SDOs (e.g., IEC/ISO JTC~3). TWAs and projects are approved by the Steering Committee to ensure alignment with NMI-Q's strategic priorities and with broader G7 objectives on quantum technologies.
\begin{figure*}[t]
    \centering %
    \safeincludegraphics[width=0.9
    \textwidth]{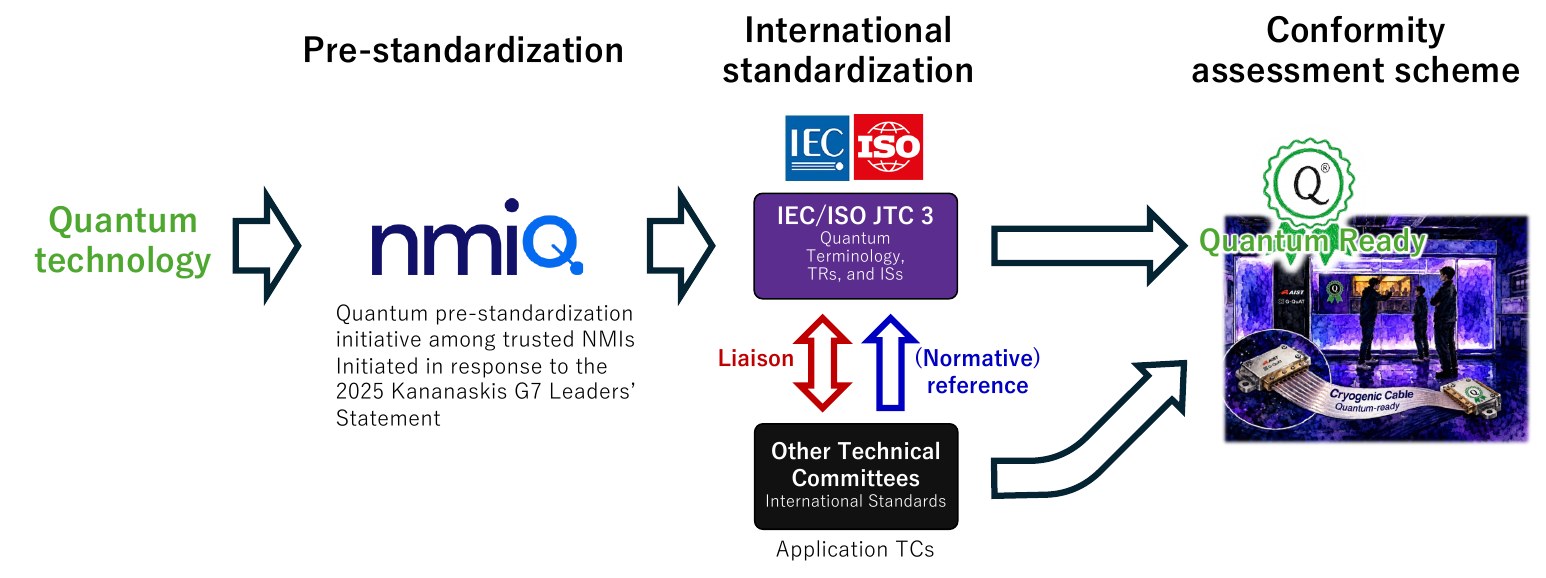}
    \caption{Standardization pathway toward a trusted quantum technology supply chain. In quantum standardization, the process should proceed from pre-standardization activities, through the development of international standardization, and ultimately to the establishment of conformity assessment schemes. When this process is completed, the necessary basis for a trusted and functioning supply chain can be considered to have been established.}
    \label{fig:Pre-standardizaiton_to_Conformity_assessment_scheme}
\end{figure*}%

NMI-Q explicitly aims to engage industry, academia, and other stakeholders not only to harmonize quantum-specific metrics---such as quantum-computing benchmarks, qubit fidelity, error rates, and quantum-sensor sensitivity---but also to address cross-cutting issues including materials characterization, security, supply-chain robustness, and quantum-resilient infrastructure. By providing objective performance benchmarks and traceable measurement frameworks, the alliance aims to reduce hype, enable fair comparisons across competing platforms, and de-risk public and private investment in the quantum sector.

In its inaugural term, NMI-Q is co-chaired by the UK and the US, reflecting their long-standing leadership in quantum metrology and standardization. Early priorities include establishing initial TWAs, launching coordinated pre-standardization projects, and defining pathways for NMI-Q outputs to feed into international standardization processes and regulatory frameworks through SDOs. As such, the establishment of NMI-Q in late~2025 represents a concrete translation of the G7's Kananaskis commitments into a durable, metrology-driven framework for trusted quantum technologies at the global level.

With the establishment of IEC/ISO JTC 3 and NMI-Q, international standardization activities for quantum technologies have entered a more active phase. In this context, it is important to clearly distinguish the respective roles of IEC/ISO JTC 3 and NMI-Q.
IEC and ISO provide international platforms for the development and consensus-based approval of standards, and IEC/ISO JTC 3, established under these organizations, is the international standardization committee responsible for quantum technologies. However, IEC/ISO JTC 3 itself is not an experimental platform for continuously and systematically conducting round-robin tests, demonstrating measurement protocols, comparing measurement results, or validating measurement methods. Such activities should be regarded as pre-standardization activities, which are positioned upstream of formal international standardization.

NMI-Q can be regarded as an international collaboration platform among national metrology institutes (NMIs) that supports such pre-standardization activities. NMI-Q is a collaborative framework centred on the NMIs of the G7 countries and Australia. Its objective is to establish the technical basis for future international standardization by developing measurement methods, test methods, best practices, interlaboratory comparisons, and reliable measurement frameworks/protocols necessary for the societal implementation of quantum technologies.

This relationship is analogous to that between VAMAS and ISO/IEC in the field of advanced materials. VAMAS was conceived in 1982 in the context of the Versailles G7 Economic Summit as an international pre-standardization framework for advanced materials \cite{ISO_VAMAS,Freiman2001VAMAS}. It has supported the development and validation of harmonized measurement and test methods, codes of practice, specifications, best practices, and standards through international collaborative projects, including interlaboratory comparisons and round-robin activities \cite{ISO_VAMAS,VAMAS_Information,BIPM_VAMAS_Workshop2016}. In this sense, NMI-Q may be regarded as a VAMAS-type pre-standardization platform for quantum technologies, or, in other words, a “quantum version of VAMAS.”

Once sufficient experimental consensus has been established within NMI-Q, its outcomes can provide technical input to the development of Technical Reports (TRs) and International Standards (ISs) within IEC/ISO JTC 3. In addition, when specific applications, components, or systems fall within the scope of existing IEC or ISO technical committees, it is desirable that application-specific ISs be developed within the relevant TCs by drawing on, and where appropriate using as normative references, the cross-cutting terminology, evaluation concepts, and measurement methods developed within IEC/ISO JTC 3.

However, the completion of ISs or TRs alone does not constitute the establishment of a quantum industrial supply chain. Standards are documents that define requirements, evaluation methods, terminology, and test methods. In order to verify in the market that actual products, components, systems, or services conform to those standards, conformity assessment schemes must also be established. Such schemes need to specify which standards are to be used, which items are to be assessed, which bodies will perform the assessment, which procedures will be followed, and what form of declaration, certification, registration, labelling, or mark will be granted.

Therefore, the societal implementation of quantum standardization should proceed through a sequence in which the reliability of measurement and test methods is first established through pre-standardization activities conducted by NMI-Q and related bodies; international standards and technical reports are then developed within IEC/ISO JTC 3 and relevant TCs; and, finally, conformity assessment schemes are established. This overall flow is illustrated in Fig.~\ref{fig:Pre-standardizaiton_to_Conformity_assessment_scheme}.

In practice, activities in NMI-Q and IEC/ISO JTC 3 are partly conducted in parallel. Nevertheless, the basic pathway is that experimental knowledge and measurement-based consensus obtained through NMI-Q should be reflected in standardization activities within IEC/ISO JTC 3 and relevant TCs.

When this sequence is completed, the institutional and technical basis for a trusted supply chain in quantum technologies can be considered to have been established.

%
\section{Quantum Computer Modalities}
\label{sec:QCModalities}
This section is \emph{not} intended as a modality-by-modality comparison of quantum platforms. Rather, it provides a minimal ``map'' of representative modalities to support the main objective of this paper: extracting cross-cutting measurement needs and metrology requirements that can underpin horizontal standardization.

A variety of hardware modalities for quantum computing have been proposed and explored. At present, superconducting qubits, silicon spin qubits, neutral-atom platforms, trapped-ion systems, and optical/photonic architectures are widely regarded as the leading contenders. Their key characteristics are summarized in Table~\ref{table:Modalities}, including the typical numbers of physical qubits demonstrated, indicative coherence times, gate speeds and fidelities, operating temperatures, and the main industrial actors active in each area.

The qualitative characteristics of each modality are as summarized here in Table~\ref{table:Modalities}; however, the numerical parameters---such as coherence times, gate times, and fidelities---vary even within a given modality because multiple qubit types exist and, furthermore, a variety of gate operations are used. As a result, it is not always possible to capture these parameters in this table using a single metric, and comparisons across modalities are not necessarily straightforward. Since providing an exhaustive listing would not align with the purpose of this paper, we note that only representative values are reported. The references cited in this section are representative examples and are not intended to be exhaustive.

\begin{table*}[t]
  \centering
  \caption{Comparison of representative quantum-computing modalities. Sources for the information in this table are provided in the in-text citations.}\label{table:Modalities}
  {\tiny
  \setlength{\tabcolsep}{1.5pt}
  \renewcommand{\arraystretch}{0.96}
  \resizebox{0.84\textwidth}{!}{%
  \begin{tabular}{@{}llllll@{}}
  \toprule
   & \tblcell{1.8cm}{\centering\textbf{Electrons}} & \tblcell{1.8cm}{\centering\textbf{Electrons}} & \tblcell{1.8cm}{\centering\textbf{Light}} & \tblcell{1.8cm}{\centering\textbf{Atoms}} & \tblcell{1.8cm}{\centering\textbf{Atoms}} \\
  \cline{2-6}
  \tblcell{1.9cm}{\textbf{Modalities}} & \tblcell{1.8cm}{Superconducting quantum computer (SCQC)} & \tblcell{1.8cm}{Silicon spin quantum computer (Si Spin QC)} &
   \tblcell{1.8cm}{Optical/photonic quantum computer (Optical/photonic QC)} &
   \tblcell{1.8cm}{Trapped-ion quantum computer (TIQC) (Ba\textsuperscript{+}, Yb\textsuperscript{+}, others)} &
   \tblcell{1.8cm}{Neutral-atom quantum computer (NAQC) (Rb, Yb, Sr, others)} \\
  \midrule
\tblcell{1.9cm}{Physical qubits/qumodes; logical qubits} &
\tblcell{1.8cm}{1121 physical qubits (IBM); logical qubits: demonstrated, small-scale} &
\tblcell{1.8cm}{12 physical qubits (Intel); logical qubits: not yet demonstrated publicly} &
\tblcell{1.8cm}{216 qumodes (Xanadu); logical qubits: N/A} &
\tblcell{1.8cm}{98 physical qubits (Quantinuum); 48 logical qubits (error-corrected)} &
\tblcell{1.8cm}{280 physical qubits (Harvard/MIT); 48 logical qubits (error-detected/encoded); \num{3000}-qubit coherent neutral-atom array/system (Harvard/MIT); more than \num{6100} qubits in a coherent neutral-atom tweezer array/system (Caltech); logical qubits: not demonstrated}\\
  \hline
  \tblcell{1.9cm}{Principle} &
  \tblcell{1.8cm}{Control and readout of superconducting qubit circuits using microwaves} &
  \tblcell{1.8cm}{Control and readout of electron spins in semiconductor quantum dots using microwaves} &
  \tblcell{1.8cm}{Information encoded in light, processed using phase shifters and beam splitters} &
  \tblcell{1.8cm}{Ions trapped in ultra-high vacuum by electromagnetic fields and controlled/read out with lasers} &
  \tblcell{1.8cm}{Neutral atoms arranged by MOT and optical tweezers, controlled/read out with lasers} \\
  \hline
  \tblcell{1.9cm}{Strengths} &
  \tblcell{1.8cm}{Fast gates; potential for integration} &
  \tblcell{1.8cm}{Very small qubits; high integration; leveraging established silicon microfabrication technology} &
  \tblcell{1.8cm}{Room-temperature operation; relatively insensitive to environment; scalability in time and space; potential for very fast operation} &
  \tblcell{1.8cm}{Very stable quantum states; essentially identical qubit characteristics} &
  \tblcell{1.8cm}{Very stable quantum states; high scalability; essentially identical qubit characteristics} \\
  \hline
  \tblcell{1.9cm}{Weaknesses} &
  \tblcell{1.8cm}{Requires about \qty{20}{\milli\kelvin} (large cooling power); complex wiring; difficult to fabricate uniform devices; qubit elements are large (inductors)} &
  \tblcell{1.8cm}{Requires about \qty{100}{\milli\kelvin}; requires sophisticated control; magnetic fields / micromagnets; difficult to fabricate uniform devices} &
  \tblcell{1.8cm}{Requires sophisticated control; photon loss; high-speed, high photon-number-resolving detector and high-squeezing-level squeezer needed for FTQC} &
  \tblcell{1.8cm}{Slow operation. UHV/XHV required} &
  \tblcell{1.8cm}{Slow operation. Ultra-high-power, highly stabilized laser and UHV/XHV required}\\
  \hline
  \tblcell{1.9cm}{Coherence time} &
  \tblcell{1.8cm}{$T_1$ typically \qtyrange{10}{100}{\micro\second}, up to $T_1 \sim \qty{1.68}{\milli\second}$ for transmon; $T_2^* \sim \qty{1.48}{\milli\second}$ for fluxonium}  &
  \tblcell{1.8cm}{$T_1 \sim \unit{\milli\second}$ to \unit{\second}, up to \qty{9.5}{\second}; $T_2^*$ from sub-\unit{\micro\second} to tens of \unit{\micro\second}, up to \qty{40.6}{\micro\second}; $T_2$ from tens of \unit{\micro\second} to \unit{\milli\second}, up to \qty{1.9}{\milli\second} (Hahn)} &
  \tblcell{1.8cm}{Infinite in principle (practically limited by propagation loss/coupling/buffering)} &
  \tblcell{1.8cm}{$T_1$: very long, not bottleneck; $T_2 > \qty{10}{\minute}$ (typically, minutes); $T_2^* = \qty{50}{\second}$, typically seconds to tens of seconds} &
  \tblcell{1.8cm}{$T_1$: very long, not bottleneck; $T_2 = \qty{12.6 \pm 0.1}{\second}$ (typically, seconds) (echo); $T_2^*$: typically \unit{\milli\second} to \unit{\second} (Ramsey)} \\
  \hline
  \tblcell{1.9cm}{Single-qubit gate (time, fidelity)} &
  \tblcell{1.8cm}{$O(\qty{10}{\nano\second})$; greater than \qty{99}{\percent}} &
  \tblcell{1.8cm}{Tens of \unit{\nano\second} to sub-\unit{\micro\second}; greater than \qty{99}{\percent}} &
  \tblcell{1.8cm}{(DV) Effective time scale is clock rate (source repetition, switching, feed-forward latency); SPAM fidelity = \qty{99.98 \pm 0.01}{\percent}, HOM visibility = \qty{99.50 \pm 0.25}{\percent} (conditional; loss not counted). (CV) Set by pulse period/loop delay/switching; \qty{6}{\mega\hertz}/\qty{36}{\micro\second} per sample (Borealis/Xanadu).} &
  \tblcell{1.8cm}{\qty{1}{\micro\second} to \qty{10}{\micro\second}; greater than \qty{99.9}{\percent}, up to about \qty{99.9999}{\percent}} &
  \tblcell{1.8cm}{Typically around \unit{\micro\second}; \qty{99.9834 \pm 0.0002}{\percent}} \\
  \hline
  \tblcell{1.9cm}{Two-qubit gate (time, fidelity)} &
  \tblcell{1.8cm}{Tens to hundreds of \unit{\nano\second}; greater than \qty{99.9}{\percent}} &
  \tblcell{1.8cm}{Tens to hundreds of \unit{\nano\second}; greater than \qty{99}{\percent}} &
  \tblcell{1.8cm}{(DV) Two-qubit fusion fidelity = \qty{99.22 \pm 0.12}{\percent} (conditional; loss not counted)} &
  \tblcell{1.8cm}{About \qtyrange{10}{200}{\micro\second}; around or above \qty{99.9}{\percent}, up to \qty{99.9 \pm 0.1}{\percent}}&
  \tblcell{1.8cm}{Typically sub-\unit{\micro\second} to a few \unit{\micro\second}; at least \qty{99}{\percent}, up to about \qty{99.5}{\percent}} \\
  \hline
  \tblcell{1.9cm}{Operating temperature of QPU} &
  \tblcell{1.8cm}{About \qty{20}{\milli\kelvin}} &
  \tblcell{1.8cm}{\qtyrange{0.1}{4}{\kelvin} (target: \qty{4}{\kelvin})} &
  \tblcell{1.8cm}{Room temperature (including homodyne measurement set-ups; however, detectors such as SNSPDs and TESs are typically operated at cryogenic temperatures, e.g., \qtyrange{0.1}{4}{\kelvin})} &
  \tblcell{1.8cm}{Room temperature (operation at \qtyrange{4}{100}{\kelvin} also studied)} &
  \tblcell{1.8cm}{Room temperature} \\
  \hline
\tblcell{1.9cm}{Companies} &
  \tblcell{1.8cm}{Fujitsu (JP); IBM (US); Google (US); Rigetti (US); AWS (US); Oxford Quantum Circuits (GB); IQM (FI); Alice \& Bob (FR); QuantWare (NL)}&
  \tblcell{1.8cm}{Hitachi (JP); Intel (US); DIRAQ (AU); Quantum Motion (GB); Equal1 (IE)} &
  \tblcell{1.8cm}{OptQC (JP); PsiQuantum (US); Xanadu (CA); Orca Computing (GB); Quandela (FR); QuiX Quantum (NL/DE)} &
  \tblcell{1.8cm}{Qubitcore (JP); IonQ (US); Quantinuum (US/GB); Universal Quantum (GB); Qudora (DE); AQT (AT)} &
  \tblcell{1.8cm}{Yaqumo (JP); QuEra (US); Infleqtion (US); Atom Computing (US); Pasqal (FR)} \\
  \bottomrule
  \end{tabular}}
  }
\end{table*}

As the table shows, superconducting qubits and silicon spin qubits represent two complementary ``electron-based'' approaches. Superconducting devices---prominently developed by IBM, Google, AWS, Rigetti, Fujitsu, and others---offer fast single- and two-qubit gates (typically \qtyrange{10}{50}{ns} for single-qubit control and \qtyrange{40}{400}{ns} for two-qubit entangling gates, depending on the gate scheme)~\cite{Krantz2019,Kjaergaard2020}. For example, single-qubit gate times around \qty{20}{ns} have been reported for transmon devices operating in a high-fidelity regime~\cite{Li2023}. State-of-the-art processors achieve single- and two-qubit gate fidelities approaching, and in selected cases exceeding, \qty{99.9}{\percent}; for instance, Ref.~\cite{Li2023} reports an average single-qubit gate error of $r_\mathrm{avg}=\num{7.42+-0.04e-5}$, corresponding to a fidelity of \qty{99.99258}{\percent}.

These advantages come at the cost of comparatively short coherence times that are often in the tens-to-hundreds of microseconds range in typical device generations~\cite{Kjaergaard2020}. In selected implementations, however, coherence can reach the millisecond regime. For example, fluxonium qubits have demonstrated $T_{2}^{*}\approx \qty{1.48+-0.13}{ms}$~\cite{Somoroff2023}. Moreover, superconducting quantum memories implemented with transmon-based circuit technology have reported millisecond-scale coherence in on-chip bosonic modes~\cite{Ganjam2024}. Finally, recent advances in 2D transmon materials and fabrication have demonstrated energy-relaxation times approaching or exceeding the millisecond level in transmon qubits~\cite{Bland2025Nature}.

Superconducting qubits moreover require operation at $\sim$\qty{20}{\milli\kelvin} in dilution refrigerators (DRs).

Silicon spin qubits, by contrast, aim to leverage the semiconductor manufacturing ecosystem. Their physical footprint is much smaller than that of superconducting qubits, which is favorable for dense integration and, ultimately, scalable fabrication. As summarized in a recent review~\cite{Burkard2023RMP}, silicon-spin coherence spans a wide range across implementations: gate-defined quantum-dot qubits routinely reach millisecond-scale coherence, while donor-spin systems in isotopically enriched $^{28}$Si can achieve coherence times extending into the seconds regime~\cite{Tyryshkin2012}. 

In terms of control performance, state-of-the-art quantum-dot devices have demonstrated single- and two-qubit gate fidelities above \qty{99.5}{\percent}~\cite{Xue2022}, and recent industry-compatible spin-qubit unit cells have reported gate fidelities exceeding \qty{99}{\percent} under foundry-relevant constraints~\cite{Steinacker2025Nature}. At present, however, silicon spin systems typically operate with fewer qubits than leading superconducting platforms, and high-fidelity control remains technically demanding. In particular, microwave delivery and magnetic-field engineering (e.g., on-chip micromagnets for EDSR control) introduce additional design complexity and packaging constraints~\cite{Burkard2023RMP}.

Optical/photonic quantum computers take yet another route: information is encoded in properties of light, such as the presence or absence of photons (discrete-variable encodings, DV) or the quadratures of continuous-variable (CV) states. Photonic platforms---exemplified by Xanadu, PsiQuantum, ORCA Computing, Quandela, and OptQC---benefit from room-temperature operation and the intrinsic immunity of photons to many environmental disturbances. In these systems, the dominant decoherence mechanism is typically optical loss rather than dephasing, so performance is often characterized via loss budgets, interference/indistinguishability metrics, and detector performance rather than by $T_{1}/T_{2}$ in the matter-qubit sense~\cite{Wang2020NatPhotonics,Giordani2023RNC}.

Recent demonstrations illustrate both the scaling potential and the appropriate quantitative figures of merit. For example, Xanadu's Borealis processor reported a programmable photonic experiment with \num{216} squeezed modes~\cite{Madsen2022Nature}. In a complementary direction toward fault-tolerant building blocks, a manufacturable photonic platform has reported primitive operation metrics including SPAM (state preparation and measurement) fidelity of \qty{99.98 \pm 0.01}{\percent}, Hong-Ou-Mandel visibility of \qty{99.50 \pm 0.25}{\percent}, two-qubit fusion fidelity of \qty{99.22 \pm 0.12}{\percent}, and chip-to-chip qubit interconnect fidelity of \qty{99.72 \pm 0.04}{\percent} (all quoted as conditional fidelities, with loss treated separately)~\cite{PsiQuantum2025Nature}. 

Measurement-based photonic schemes using large cluster states offer an elegant route to fault tolerance, but they place stringent demands on low-loss integrated photonic circuits, high-squeezing sources with low excess noise, and high-efficiency photon-number-resolving detectors such as transition-edge sensors (TESs)~\cite{Bourassa2021Quantum,Wang2020NatPhotonics}. In particular, increasing the maximum count rate while preserving photon-number resolution is important for faster measurements at higher photon flux, and remains an active area of research and engineering.

Trapped-ion and neutral-atom systems encode qubits in internal states of individual ions or atoms confined in electromagnetic or optical traps. These atomic platforms---pursued by companies such as Quantinuum, IonQ, AQT, Infleqtion, Pasqal, QuEra, Yaqumo, and Atom Computing---offer intrinsically long coherence, together with high-fidelity control and favorable connectivity (often all-to-all for trapped ions and long-range or programmable connectivity for neutral atoms)~\cite{Bruzewicz2019APR,Saffman2010RMP,Henriet2020Quantum}. In trapped-ion systems, record single-qubit performance includes coherence times exceeding 10 minutes under dynamical decoupling and single-qubit gate fidelities as high as 99.9999\%~\cite{Wang2017NatPhotonics,Harty2014PRL}. In large neutral-atom tweezer arrays, hyperfine-qubit coherence can likewise reach the multi-second regime; for example, \(T_{2}=12.6(1)\,\mathrm{s}\) has been reported under dynamical decoupling, alongside a global single-qubit randomized-benchmarking fidelity of \(99.9834(2)\%\)~\cite{Manetsch2025}. Two-qubit entangling operations are also rapidly improving. For trapped ions, two-qubit gate fidelities of 99.9(1)\% have been demonstrated for hyperfine qubits~\cite{Ballance2016PRL}, while in neutral-atom systems, a CZ-gate fidelity of 99.5\% on up to 60 qubits in parallel has been reported~\cite{Evered2023Nature}.

The main drawbacks of atomic platforms are slower entangling-gate speeds---typically in the microsecond to hundreds-of-microseconds regime---together with stringent requirements on lasers, optical alignment, and ultra-high-vacuum infrastructure, and non-trivial scaling as the number of trapped particles and optical channels grows~\cite{Bruzewicz2019APR,Saffman2010RMP,Saffman2016JPB,Henriet2020Quantum}. Neutral-atom machines, which use optical tweezers or optical lattices to form large arrays, are particularly promising for scaling to hundreds or thousands of physical qubits (and beyond), but still require advances in optical control, calibration and automation, and error-correction integration to reach fully fault-tolerant operation~\cite{Henriet2020Quantum,Saffman2016JPB}.

Overall, Table~\ref{table:Modalities} underlines that no single modality is universally superior: fast but relatively short-lived qubits (superconducting and semiconductor spins) coexist with slower but extremely stable atomic and photonic qubits. Each technology occupies a different point in the landscape of gate speed, coherence, connectivity, temperature, and manufacturability, and each is being driven forward by a distinct industrial and academic ecosystem. It is therefore plausible that multiple hardware platforms will coexist for the foreseeable future, with the most suitable choice depending on the target application, the level of fault tolerance required, and broader system-level constraints such as cooling infrastructure and supply-chain maturity.

Table~~\ref{table:Modalities} should be read as a \emph{navigation aid}, not a scorecard. Its role is to make explicit the recurring measurement needs shared across modalities, thereby motivating cross-cutting metrology infrastructure and standardization pathways discussed in the subsequent sections.

\section{Cross-cutting development and standardization of hardware components across modalities}\label{sec:crosscutting}

\begin{table*}[t]
\centering
\caption{Cross-modality common subsystems and enabling technologies.}
\label{tab:cross_modality_subsystems}
{\scriptsize
\setlength{\tabcolsep}{3pt}
\renewcommand{\arraystretch}{1.12}
\begin{tabular}{@{}lccccc@{}}
\toprule
 & \textbf{SCQC} & \textbf{Si Spin QC} & \textbf{Optical/Photonic QC} & \textbf{TIQC} & \textbf{NAQC} \\
\midrule
\tblcell{4.6cm}{Cryogenics (4\,K and above; GM/PTR-based)} & & $\checkmark$ for hot spin qubit & $\checkmark$ for SNSPD & $\checkmark$ & \\
\tblcell{4.6cm}{Cryogenics (dilution refrigerator)} & $\checkmark$ & $\checkmark$ & $\checkmark$ for TES & & \\
\tblcell{4.6cm}{(Cryogenic) QPU packaging} & $\checkmark$ & $\checkmark$ & & $\checkmark$ & \\
\tblcell{4.6cm}{(Cryogenic) Wiring} & $\checkmark$ & $\checkmark$ & & $\checkmark$ & \\
\tblcell{4.6cm}{Laser} & & & $\checkmark$ & $\checkmark$ & $\checkmark$ \\
\tblcell{4.6cm}{TES, SNSPD} & & & $\checkmark$ & & \\
\tblcell{4.6cm}{SPAD} & & & $\checkmark$ & $\checkmark$ & $\checkmark$ \\
\tblcell{4.6cm}{Photon-number-resolving CMOS camera} & & & & $\checkmark$ & $\checkmark$ \\
\tblcell{4.6cm}{Non-magnetic material} & $\checkmark$ & $\checkmark$ & $\checkmark$ for TES and SNSPD & $\checkmark$ & $\checkmark$ \\
\bottomrule
\end{tabular}
}
\end{table*}

This section identifies cross-cutting opportunities for metrology-led standardization by focusing on enabling component technologies that recur across quantum-computing modalities.
Table~\ref{tab:cross_modality_subsystems} provides a compact overview of such subsystems that are amenable to horizontal development and standardization.
Read alongside Table~\ref{table:Modalities}, it shows that multiple modalities depend on closely related underlying technologies, which motivates shared terminology, metrics, and test methods at the component level as a practical first step in ``charting a pathway'' to standardization.

For certain component technologies, it is advantageous to horizontally transfer (cross-leverage) techniques from more mature modalities, enabling more efficient development of higher-performance systems on a cross-modality, horizontal platform. This implies that cross-cutting (cross-modality or horizontal) development of hardware components is not only feasible but also highly desirable: by co-developing shared building blocks, we can deliver hardware that is both higher performing and more cost-effective.

Such an approach can also broaden participation in the emerging quantum industry by enabling a wider range of companies and industrial sectors to contribute. It is likewise beneficial for standardization and for establishing robust and diversified supply chains. In this section, we highlight several representative items and describe the quantum-computing methods and modalities in which they are employed.

\subsection{Cryogenics}
Cryogenic infrastructure---most notably dilution refrigerators (DRs)---can be developed within a largely common framework for both superconducting and silicon spin quantum computers (SCQCs and Si spin QCs), because both platforms rely on high cooling power at millikelvin temperatures. Other modalities, such as optical/photonic quantum computers, are also expected to require cryogenic systems, since state-of-the-art photon-number-resolving detectors, including superconducting transition-edge sensors (TESs), generally require sub-kelvin operation.

A key difference, however, is that superconducting and silicon-spin platforms require the DR cooling power to scale with system size, whereas optical/photonic quantum computers typically require cooling primarily for photon-number-resolving detector arrays. As a result, many photonic approaches may be accommodated by continued miniaturization and maintenance-free operation of existing cryogenic systems, rather than by a proportional increase in refrigeration capacity as the overall system scales.

Trapped-ion quantum computers (TIQCs) are increasingly using trap chips cooled to around \qtyrange{4}{100}{\kelvin}, implying a growing need for low-vibration GM (Gifford-McMahon) and PT (pulse-tube) refrigerator-based cooling systems. As the qubit count scales---i.e., as the number and size of trap chips increase---and as the heat load from devices co-packaged with the trap chips rises, the required cryogenic capacity of pulse-tube refrigerators (PTRs) and Gifford-McMahon refrigerators (GMRs) will increase accordingly. Consequently, continued development of cryogenic refrigeration technologies will be essential. In particular, scaling to large systems is likely to require substantial investment when operation below $\sim$\qty{10}{\kelvin} is required.

For temperature regimes at and above $\sim$\qty{20}{\kelvin}, in addition to GMRs and PTRs, cryogenic refrigeration technologies developed for liquid-hydrogen (L-H$_2$) production and storage---such as \qty{20}{\kelvin}-class Brayton/Claude-based cryogenic processes and recondensation/reliquefaction schemes---may be applicable, depending on the required cooling capacity and operating constraints. Moreover, when large-scale refrigeration (e.g., \unit{kW}-class) is required below $\sim$\qty{10}{\kelvin}, system-level technologies established in helium liquefaction and refrigeration plants, including multi-stage refrigeration architectures, high-effectiveness heat exchangers, and turbine expanders, may be transferable or at least provide a useful design basis.

These cryogenic refrigeration technologies can also be applied to DR systems, particularly to provide high cooling capacity at the $\sim$\qty{4}{\kelvin} precooling stage. In this sense, they can form part of the technological basis for scaling up large-capacity DR platforms for SCQCs and Si spin QCs.

Si spin QCs already operate at somewhat higher temperatures than SCQCs, on the order of \qty{100}{\milli\kelvin}, and there is active work toward raising the operating temperature further. This work targets ``hot'' silicon spin qubits, aiming for operation above \qty{1}{\kelvin}~\cite{Yang2020_Above1K,Petit2020_HotSi,Huang2024_Above1K} and pursuing an ambitious extension toward \qty{4}{\kelvin}-class qubit operation~\cite{Camenzind2022_Above4K}, motivated by relaxed cryogenic requirements and tighter integration with control electronics.

If achieved, this would provide a substantial advantage in required cooling power and overall system complexity, since such operation could be realized without a DR. For the foreseeable future, however, Si spin QCs will still demand cooling capabilities comparable to those of SCQCs. As a result, these two modalities will continue to rely on essentially the same class of DR platforms, and their cryogenic hardware is likely to co-evolve on a common technological base. It is therefore natural to expect that the associated supply chains and standardization activities for cryogenic infrastructure will also develop in a closely aligned manner.

Standards for low-temperature measurement are well established, and a wide range of thermometers is routinely used in DRs, GMRs, and PTRs. Since cryogenic thermometry technologies and their associated standards are already mature, we do not discuss them further here.

\subsection{Cryogenic interconnects and RF components}\label{subsection:cables}
IEC TC~46~\cite{IEC-TC46}, ``Cables, wires, waveguides, RF connectors, RF and microwave passive components and accessories,'' provides a mature standardization framework for RF cables and related components. Measurement standards in this domain are also well established, and a broad range of RF components has been commercialized and standardized. However, existing products and standards largely target operation near room temperature: cryogenic-rated products remain scarce, and dedicated standards for low-temperature use are essentially absent.

Because cryogenic RF components are broadly enabling across multiple quantum-computing modalities, it is imperative to advance both (i) product development of compact, high-density, wideband (and/or band-specific) components and (ii) fit-for-purpose standardization and specification work in parallel. In particular, there is an urgent need to address high-density connectors, high-density flat (ribbon) cables, attenuators, filters, amplifiers, and isolators spanning roughly the MHz to $\sim$\qty{20}{GHz} range. Lower-frequency bands are likely to be broadly used across SCQCs, Si spin QCs, and TIQCs, whereas the higher-frequency regime will be used primarily by SCQCs.

As the technology matures and scalability improves, modality-specific specifications will be scrutinized with increasing precision. In parallel, cross-cutting (horizontal) standards should be developed across modalities, and, where necessary, modality-specific (vertical) standards should be established.

\subsection{Chip carriers and packaging}\label{subsection:CC}
In advanced silicon process technologies, ultra-dense, heterogeneous packaging has been developed at room temperature for conventional PCs and high-performance computers (HPCs; e.g., large GPU clusters and supercomputers). For example, multiple small dies (``chiplets'') can be integrated or co-packaged within a single package, with high-density die-to-die interconnects via an interposer (2.5D/3D integration) to deliver higher bandwidth and better system-level performance.
A further trend is to relocate power delivery to the backside of the die (backside power delivery), freeing front-side routing resources for signals and improving power/signal integrity---an approach now being productized for leading nodes~\cite{HIR2021,intel}.

In parallel, the industry is moving toward co-packaged optics, bringing optical I/O (often based on silicon photonics (Si/SiN platforms)) into close proximity with switching/compute ASICs to reduce electrical I/O power and extend bandwidth scaling. These packaging directions require advanced high-speed electrical/RF design (signal integrity, power integrity) as well as rigorous thermal co-design at the package and system levels.

To relieve the interconnect explosion at the quantum processing unit (QPU) package boundary, it is increasingly plausible that heterogeneous co-packaging---already mainstream in room-temperature compute (chipletization, 2.5D/3D integration, and tightly coupled ``chipset'' partitioning of functions as described above)---will become the default architectural pattern for cryogenic QPUs as well.

The motivation is straightforward: as systems push toward large-scale operation, the scaling bottleneck shifts from qubit devices to the surrounding infrastructure (wiring density, thermal load, refrigerator geometry, and the sheer footprint of control/readout resources). Recent analyses of cryoelectronics integration emphasize that fault-tolerant superconducting systems, in particular, demand a commensurate scaling of the classical stack, making integration strategy a first-order design variable rather than an afterthought~\cite{kawabata2026}. In parallel, the semiconductor packaging community has matured chip-package-board co-design and multiphysics workflows (electrical-thermal-mechanical signoff, reliability modeling, and package-level simulation driven by electronic design automation (EDA)), which can be repurposed with comparatively little friction for cryogenic system partitioning and early design-space exploration~\cite{HIR2021}. Demonstrators explicitly targeting 3D co-integration of quantum devices with Cryo-CMOS control chips (e.g., CEA-Leti's QuIC3 direction~\cite{LETI2022}) reinforce that the industry is already testing the ``chiplet-like'' playbook in the sub-kelvin regime~\cite{lapedus2021}.

The key complication is not the absence of simulation engines, but the lack of validated cryogenic material data for package-relevant stacks. Advanced packaging models become fragile when basic inputs---thermal conductivity, specific heat, and thermal expansion (CTE; coefficient of thermal expansion)---are poorly known or strongly process-dependent at cryogenic temperatures. Without a critically curated property database, electro-thermal and thermo-mechanical simulation cannot be trusted for stress hot-spots, warpage, delamination risk, or thermal-anchoring design, especially when mixing silicon, organic build-up layers/underfills/adhesives, LTCC/HTCC (low-/high-temperature co-fired ceramics) substrates integrating multilayer interconnects, diverse ceramic compositions, and related insulators. Also, elastic modulus, creep/fatigue behavior, interfacial adhesion, and fracture toughness will become increasingly important as packages increase in size, integration density, and overall complexity in the future.

Existing cryogenic property efforts (notably NIST's compilations and reference lists spanning thermal and mechanical properties in roughly \qtyrange{4}{300}{\kelvin} for common cryogenic materials) provide an essential starting point, but they do not fully cover modern packaging material systems, vendor/process variations, or temperatures down to the sub-kelvin regime that dominate real assemblies~\cite{Bradley2013,NIST-LIST}. Therefore, a disciplined measurement program---covering silicon, polymers/organics, ceramics, and crystals used in interposers, substrates, underfills, and structural dielectrics---should be treated as enabling infrastructure.

In addition to cryogenic RF/DC electrical, thermal, and thermomechanical properties, low-outgassing performance is a critical materials requirement for vacuum-based platforms. Low-outgassing materials engineered to meet UHV/XHV (ultra-high vacuum/extreme-high vacuum) requirements---reducing outgassing to the lowest practicable levels---are expected to become increasingly important for TIQC platforms, particularly for packaging materials, and will also benefit other modalities such as NAQCs.

If the resulting datasets are curated into an interoperable database and aligned with standard test methods, they become broadly reusable across essentially all cryogenic quantum-computing modalities (SCQC, Si spin QC, TIQC, photonic/optical QC, and diamond-based platforms), making this both a cross-cutting technical contribution and a credible standardization target.

\subsection{Optical and photonic subsystems}\label{subsection:Optics}
Across several leading platforms, optical subsystems constitute shared infrastructure that can effectively set the upper bound on qubit count and overall computing capability. Neutral-atom systems, in particular, have demonstrated rapid growth in array sizes and site counts, which in turn tightens requirements on the trapping and control light (power, relative intensity noise (RIN), frequency/phase noise, beam quality, short-/long-term stability, and control bandwidth), as well as on beam-shaping and addressing hardware (e.g., spatial light modulators (SLMs)) and the readout chain (e.g., objective lenses and detectors). These parameters map directly onto atom loss, gate speed, and gate error rates as the system scales. Recent large optical-tweezer arrays, for example, underscore that platform progress is now strongly coupled to the robustness and uniformity of the optical layer~\cite{Manetsch2025}.

In parallel, optical/photonic quantum computing---much like neutral-atom platforms---is converging on architectures in which laser stability and coherence are engineering constraints rather than lab-grade ``nice-to-have'' features. Of course, this modality requires high-performance, high-squeezing-level squeezers; more broadly, its end-to-end performance hinges on optical sources whose frequency (or phase) noise, RIN, and drift are sufficiently small and repeatable across many channels and long runtimes. Whether photonic qubits are generated and processed on-chip or in hybrid modules, these requirements become stricter as designs move toward larger, more integrated fault-tolerant roadmaps. In practice, this is pushing laser development toward cavity-referenced stabilization, optical-frequency-comb-referenced locking and phase-coherent synchronization across multiple lasers, tighter environmental hardening, and better-specified performance envelopes at relevant wavelengths and power levels (rather than ``best-effort'' performance on a single optical table).

The same ``shared infrastructure'' logic applies to detectors and wavefront/beam-shaping components. Single-photon detection technologies---e.g., superconducting nanowire single-photon detectors (SNSPDs)---and photon-number-resolving (PNR) devices---e.g., transition-edge sensors (TESs)---are being pulled simultaneously by quantum communication, sensing/metrology (including minimally invasive biomedical measurement and imaging), and computing, because they sit on critical readout and verification paths in all of these areas. Recent reviews explicitly frame single-photon detectors and PNR devices as enabling components spanning sensing, communications (e.g., quantum key distribution (QKD)), and other quantum photonics use cases---exactly the kind of overlap that supports common qualification methods and cross-platform specifications. For photon-number resolution in particular, AIST, NIST~\cite{NIST-IR-8486}, and others have documented detector development and performance metrics that are broadly reusable across quantum applications, and the research community continues to refine how PNR capability should be defined and evaluated in practice.

Given this landscape, it is strategically cleaner to treat quantum-optical components as a family of reusable building blocks and to pursue standardization in a platform-agnostic manner by default: define common performance descriptors, test conditions, uncertainty reporting, and interoperability interfaces first, then add modality-specific annexes only where physics or architecture truly forces divergence. This approach also aligns with the remit of IEC/ISO JTC~3, whose scope explicitly covers quantum sources and detectors alongside computing, communications, and metrology---providing a natural home for cross-platform work on optical devices and their measurement methods.

\subsection{Magnetic properties of materials and components}
\begin{figure*}[t]
    \centering %
    \safeincludegraphics[width=0.6
    \textwidth]{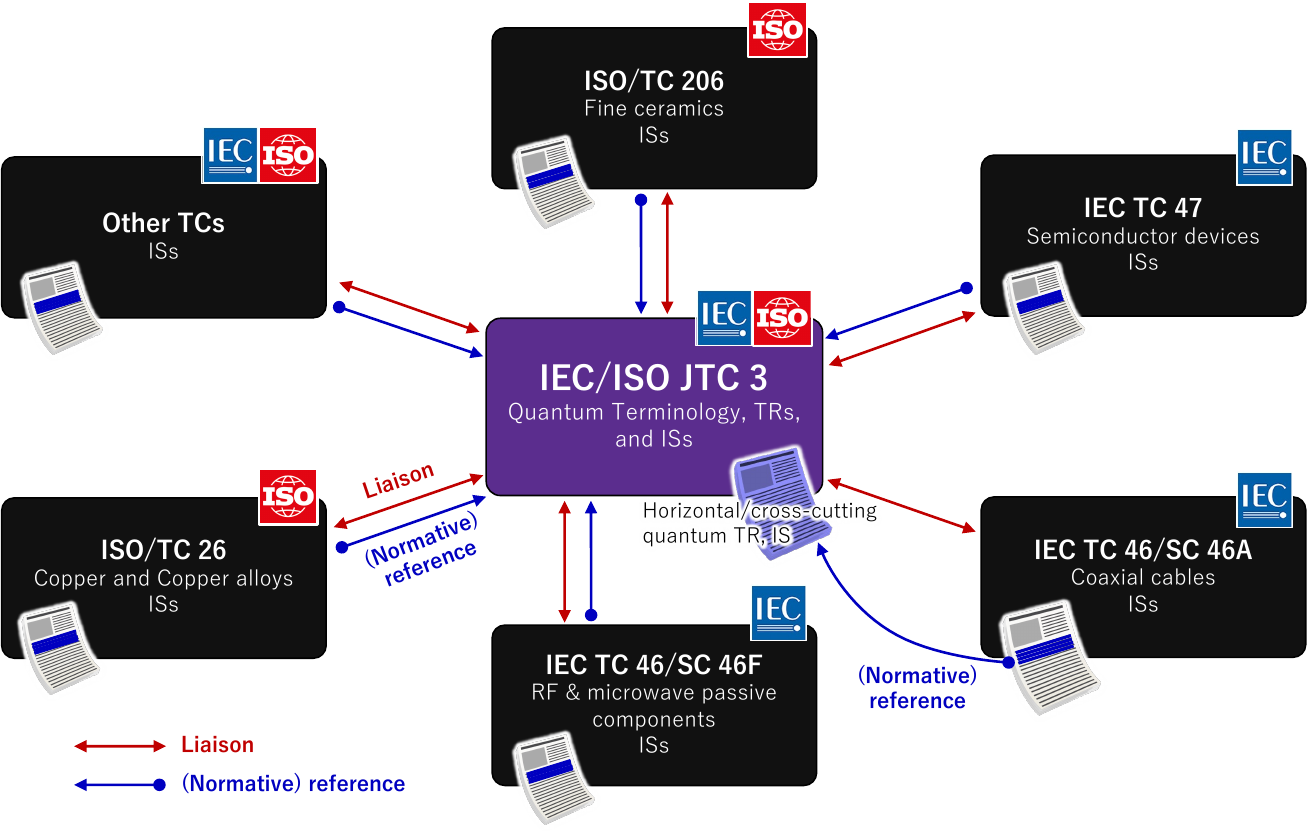}
    \caption{A schematic of the proposed collaboration framework for non-magnetic components between IEC/ISO JTC 3 and technical committees (TCs) of ISO, IEC, and other standards development organizations (SDOs). Specific TCs develop International Standards (ISs) for application- or sector-specific evaluation and test methods, while IEC/ISO JTC 3 develops Technical Reports (TRs) and International Standards (ISs) that provide cross-cutting terminology, definitions, and evaluation methodologies for the magnetic properties of materials and components spanning multiple TCs. TCs may establish liaisons with IEC/ISO JTC 3 and develop ISs by drawing on the TRs developed within IEC/ISO JTC 3 and, where appropriate, normatively referencing the relevant ISs developed within IEC/ISO JTC 3. Where relevant TCs lack the technical capacity to develop an IS, IEC/ISO JTC 3 may itself initiate and develop such an IS, as appropriate. The TCs for which liaisons have already been established as of June 2026 are listed in Table~\ref{tab:Liaisons_2026Feb}. This framework helps ensure consistency and avoid discrepancies across TC activities.}
    \label{fig:Non-mag}
\end{figure*}
%
Qubits are generally highly sensitive to external magnetic and electric fields; accordingly, the magnetic properties of nearby components and materials can degrade device performance, for example by shortening coherence times. In addition, superconducting circuit technologies such as RSFQ (rapid single-flux quantum) and AQFP (adiabatic quantum-flux-parametron) devices can also be adversely affected by nearby magnetic materials, potentially leading to malfunction, spurious switching, or even loss of operation. Rigorous control of the magnetic environment is therefore essential in the vicinity of both qubits and superconducting circuitry.

In practice, magnetic shielding is widely employed to reduce the residual magnetic flux density inside the shield to roughly
$10^{-3}$ of the ambient level---typically on the order of nT---relative to the geomagnetic field ($\sim$\qty{50}{\micro\tesla}).
Consequently, for components placed inside the magnetic shield---especially those in the immediate vicinity of qubits and superconducting circuits---materials must be selected with particular care to ensure negligible magnetism. Magnetic shields are typically implemented using superconducting materials (e.g., Al ($T_{\mathrm{c}} \approx \qty{1.2}{\kelvin}$) or Nb ($T_{\mathrm{c}} \approx \qty{9.2}{\kelvin}$)) that exploit perfect diamagnetism (the Meissner effect) and/or cryogenic high-permeability alloys (relative permeability on the order of $10^{5}$), which are commercially available as cryogenic ``$\mu$-metal'' (e.g., under trade names such as Cryoperm and A4K). These high-permeability alloys are high-Ni soft-magnetic materials, typically comprising approximately \qtyrange{75}{80}{\percent} Ni, \qtyrange{15}{20}{\percent} Fe, a few percent Mo, and trace amounts of Cu.

Because such low residual-field conditions are required, magnetically compliant design practices are applied throughout the shielded volume that houses superconducting qubit devices. These practices include avoiding magnetic materials in structural parts, screws/nuts/fasteners, wiring, connectors, and plating stacks. For example, at the coldest stage of a dilution refrigerator, oxygen-free copper (OFC)---generally regarded as non-magnetic and offering high thermal conductivity---is widely used. Its surface is often gold-plated to prevent oxidation and to minimize and stabilize electrical and thermal contact resistance.

However, ``non-magnetic'' in practice is not determined solely by bulk material choice: fabrication and handling can introduce magnetic contamination (e.g., transfer of ferromagnetic debris from steel tools), and the resulting surface magnetism can dominate the intrinsic susceptibility of otherwise benign materials. Accordingly, minimizing magnetic impurities requires an end-to-end approach, including (i) preference for intrinsically low-susceptibility materials where feasible, (ii) avoidance of alloys with higher or highly variable magnetism, (iii) machining and assembly practices that limit contact with ferromagnetic tooling and fixtures, (iv) post-processing/cleaning steps to remove surface contamination when appropriate (e.g., brief dilute-acid cleaning under controlled conditions), and (v) incoming inspection and screening to manage batch-to-batch variability from suppliers and processes\cite{Sunderland2009}.

In conventional (room-temperature) hardware, gold plating is frequently applied on top of a Ni underlayer. The Ni underlayer acts as a diffusion barrier and improves adhesion and smoothness of the Au film, thereby enhancing reliability for soldering and wire bonding. However, Ni plating can be magnetic and is therefore often avoided in the packaging of superconducting qubit devices. As a result, copper parts located close to qubit chips frequently employ direct Au plating on Cu without a Ni underlayer. Stainless steel can also exhibit magnetism depending on alloy composition and processing; thus, its use in components is often restricted near superconducting qubits.

It is worth noting that the practical trade-off differs between room-temperature and cryogenic environments. At room temperature, direct Au plating on Cu can lead to Au--Cu interdiffusion, which may increase contact resistance and promote delamination. By contrast, under cryogenic conditions, Au diffusion into Cu is strongly suppressed, and maintenance operations such as component replacement or repeated connector mating are typically infrequent. Therefore, degradation mechanisms that are critical at room temperature tend to be less pronounced at cryogenic stages.

Overall, quantum computers often impose modality- and architecture-specific design constraints related to magnetism. Although OFC is commonly treated as non-magnetic, weak magnetic signals have been reported at low temperature, likely originating from trace magnetic impurities~\cite{Bowers1956,Pal2022}. In many cases, the resulting magnetization is small enough to be neglected; however, as systems scale to larger qubit counts and margins on allowable magnetic field and magnetization tighten, such effects may become non-negligible.

Quantitatively defining and evaluating (i) tolerable magnetic-field levels and (ii) allowable magnetism of materials placed near superconducting qubit devices (e.g., magnetization and residual flux) is inherently challenging. Small-scale systems have therefore traditionally been designed conservatively, with the aim of eliminating magnetic contributions as much as possible. Looking toward large-scale implementations, however, it will be necessary to establish evaluation criteria and standardized test methods for magnetic properties. 

Related standardization and specification activities are already being discussed within organizations such as IEC/ISO JTC~3. Importantly, magnetic-property evaluation should be formulated in a temperature-agnostic framework that can be applied across multiple modalities. Likewise, magnetic shielding and ``non-magnetic'' materials should be categorized by use case and required performance level, and then developed, standardized, and specified in a systematic manner.

Because the relevant material space is broad---spanning structural metals, ceramics, packaging materials, wiring and interconnect materials, and associated surface finishes---these efforts should not be confined to a single modality. Rather, cross-cutting, cross-modality development and standardization will be required.

A schematic is shown in Fig.~\ref{fig:Non-mag}, illustrating that IEC/ISO JTC 3 should address cross-cutting standardization, while specific TCs are expected to develop TC-specific standards for non-magnetic materials and components. TC-specific ISs may draw on the TRs and terminology developed by IEC/ISO JTC 3 and, where appropriate, normatively reference the relevant ISs developed within IEC/ISO JTC 3.

The collaboration model between IEC/ISO JTC 3 and specific TCs described here can also be applied to the technologies discussed in the preceding subsections.

\subsection{Characterization of color centers}
\begin{figure*}[t]
    \centering %
    \safeincludegraphics[width=0.8\textwidth]{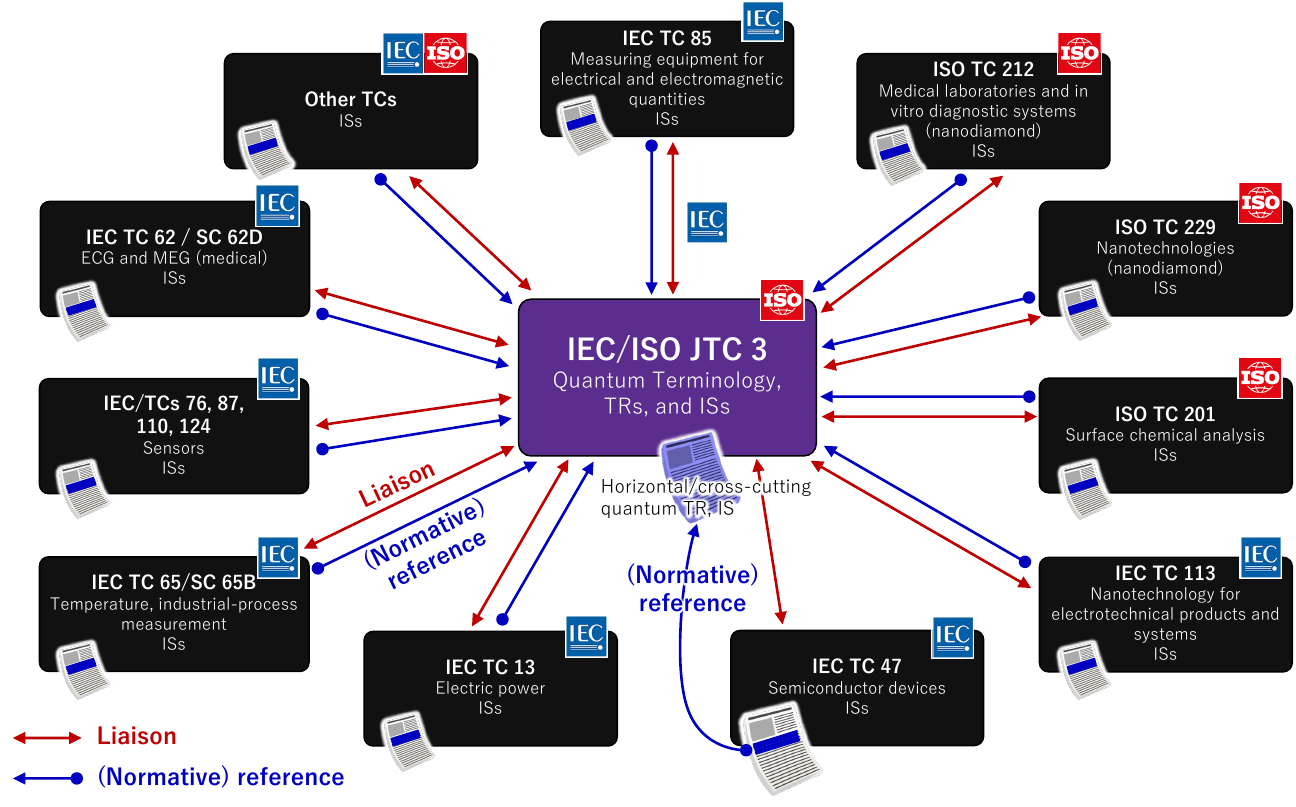}
    \caption{A schematic of the proposed collaboration framework for color-center characterization between IEC/ISO JTC 3 and technical committees (TCs) of ISO, IEC, and other standards development organizations (SDOs). Specific TCs develop International Standards (ISs) for application- or sector-specific evaluation and test methods, while IEC/ISO JTC 3 develops Technical Reports (TRs) and International Standards (ISs) that provide cross-cutting terminology, definitions, and evaluation methodologies for diamond NV centers across multiple TCs. TCs may establish liaisons with IEC/ISO JTC 3 and develop ISs by drawing on the TRs developed within IEC/ISO JTC 3 and, where appropriate, normatively referencing the relevant ISs developed within IEC/ISO JTC 3. Where relevant TCs lack the technical capacity to develop an IS, IEC/ISO JTC 3 may itself initiate and develop such an IS, as appropriate. The TCs for which liaisons have already been established as of June 2026 are listed in Table~\ref{tab:Liaisons_2026Feb}. For medical applications such as electrocardiography (ECG) and magnetoencephalography (MEG), TC-specific standardization may involve IEC TC 62 and its relevant subcommittees, including SC 62D for electromedical equipment. This framework helps ensure consistency and avoid discrepancies across TC activities.}
    \label{fig:DiamondNVC}
\end{figure*}

Although this paper primarily addresses standardization pathways for quantum-computing hardware, the underlying metrology-driven logic extends directly to quantum sensing, where cross-cutting development and standardization of components and materials can often be transferred across application domains. Since a comprehensive survey of sensing modalities is beyond the scope of the present manuscript, diamond nitrogen-vacancy (NV) centers are used here as a representative case study and are discussed in somewhat greater detail, with the aim of providing a reusable reference for future standardization discussions across other quantum-sensing platforms. Notably, diamond NV centers are also viable qubit platforms; accordingly, parts of the discussion below are relevant from both sensing and computing-hardware perspectives.

Diamond NV centers enable quantum-state preparation, coherent control, and readout, thereby forming a practical platform for quantum-enabled sensing. Because the NV spin resonance is perturbed by multiple external stimuli, NV-diamond sensors are intrinsically multimodal: magnetic~\cite{Taylor2008,Maertz2010} and electric fields~\cite{Dolde2011} are encoded as resonance shifts and splittings, temperature~\cite{Acosta2010} can be inferred via temperature-dependent spin-Hamiltonian parameters (e.g., the zero-field splitting $D(T)$), and strain (and pressure via strain-coupled perturbations) can be accessed through lattice-coupled perturbations~\cite{Hsieh2019}. These capabilities motivate broad deployment across biomedicine, environmental monitoring, and industrial metrology, often under demanding boundary conditions such as wide frequency bandwidths (including DC, depending on the measurement protocol) and wide operating temperature ranges. Consequently, NV-based sensing spans multiple ``measurement domains'' that are frequently governed by different SDOs and TCs.

This breadth of application domains creates structural challenges for standardization. If evaluation and qualification are pursued independently within application-specific TCs (i.e., vertical standardization), the same NV material or device can be forced into multiple, partially overlapping standards, each defining its own metrics, test conditions, and reporting formats. Such fragmentation is likely to introduce redundancy and, more importantly, inconsistencies in definitions and methods across international standards. It can also impose duplicated qualification workloads on materials suppliers and sensor manufacturers, while reducing the portability of characterization data across markets and undermining comparability of performance claims for end users.

A cross-cutting or horizontal standardization strategy is therefore essential. The key observation is that, despite diverse applications, NV sensing shares a common physical measurement basis---spin-resonance spectroscopy and readout---together with a common set of core material/device properties. In practice, two closely related primitives dominate: electron spin resonance (ESR) and optically detected magnetic resonance (ODMR).

Here, the NV spin resonance is interrogated in the microwave domain. In this paper, ODMR denotes optical readout via fluorescence contrast, whereas ESR denotes electrically/microwave-detected schemes (e.g., resonator-based reflectometry), unless stated otherwise.

Since both ESR and ODMR interrogate the same underlying spin-resonance physics, they can be integrated into a unified, application-agnostic characterization backbone. In such a framework, application-specific (vertical) standards do not re-define NV characterization from first principles; instead, they draw on the horizontal framework and, where appropriate, normatively reference it, adding only domain-specific requirements such as target range, traceability and calibration route, packaging constraints, or sector-specific safety requirements.

For interoperability, a horizontal ESR/ODMR framework should specify: (i) the parameters and figures of merit to be reported, (ii) measurement methods and procedures, and (iii) measurement conditions and reporting rules (including uncertainty statements where appropriate). The parameter set should include resonance observables and derived quantities that remain meaningful across applications and readout modalities: resonance frequencies, linewidth and the adopted lineshape model, contrast (ODMR fluorescence contrast, or an electrically/microwave-detected readout amplitude with a specified normalization, e.g., absorption/dispersion/reflectometry), and the dependence of these observables on optical and microwave driving conditions. Because coherence and relaxation govern sensitivity and bandwidth, coherence-related metrics---
$T_{1}$,
$T_{2}^{*}$,
and
$T_{2}$
---should be defined as standardized reportables, extracted under explicitly specified protocols. Sensitivity-relevant figures of merit should further be defined to separate intrinsic material/device contributions from application-dependent electronics, for example through standardized definitions based on resonance slope, noise spectral density under stated detection bandwidth, and duty cycle.

Methodologically, the framework should encompass both continuous-wave (CW) and pulsed protocols under ESR and ODMR. CW-ODMR and CW-ESR provide rapid screening and baseline resonance characterization under defined optical/microwave power and detection settings. Pulsed protocols are required for quantitative coherence assessment and for performance-relevant metrics; accordingly, standardized sequences such as Rabi, Ramsey (for $T_{2}^{*}$), and Hahn-echo (for $T_{2}$) should be included, with unambiguous definitions of timing conventions, phase cycling (if used), fitting models, acceptance criteria, and reporting of fit quality. Optional annex methods (e.g., dynamical decoupling) may be added provided their analysis pipelines are specified reproducibly.

Equally important is the control and disclosure of measurement conditions. ESR/ODMR observables are strongly dependent on experimental settings; thus, the framework should require explicit reporting of optical wavelength and intensity at the sample (when applicable), microwave frequency span and power at a defined reference plane, coupling structure (resonator/antenna/stripline geometry or a standardized proxy), bias-field magnitude and orientation relative to crystal axes, and the magnetic environment (including shielding description or residual-field characterization). Temperature should be measured and reported with sufficient resolution because it affects both intrinsic NV parameters and apparatus behavior. The objective is not to enforce identical setups, but to ensure that reported results are interpretable and reproducible across laboratories.

Finally, a viable horizontal framework must accommodate the diversity of NV material forms---single-crystal, polycrystalline, and nanodiamond---while preserving a minimal common set of evaluation items that enables cross-implementation comparison. Form-specific modules (e.g., orientation/strain reporting for single crystals; surface termination and aggregation descriptors for nanodiamonds) can be appended without compromising the common backbone.

In summary, NV-center-based sensing illustrates why horizontal standardization is a prerequisite for scalable, interoperable quantum-sensor markets. Establishing a unified ESR/ODMR-based characterization backbone---covering parameters, procedures, and reporting conditions across CW and pulsed protocols and across material forms---would allow vertical application standards to draw on, and where appropriate normatively reference, a common foundation rather than duplicating or diverging in core definitions. This structure reduces redundant qualification effort, improves reproducibility and comparability, and aligns naturally with the mandate of cross-domain working groups (e.g., IEC/ISO JTC 3 WG10: Quantum Sensors) to organize interoperable sensing standards that complement application-oriented TCs.

A schematic is shown in Fig.~\ref{fig:DiamondNVC}, illustrating that IEC/ISO JTC 3 should address cross-cutting standardization, while specific technical committees are expected to develop application-specific standards for diamond NV center materials and sensing technologies.



\section{Summary}\label{sec:summary}
Quantum computing is increasingly consolidating around five leading modalities---superconducting qubits, silicon spin qubits, neutral-atom platforms, trapped-ion systems, and optical/photonic architectures---while still leaving room for more challenging approaches such as topological qubits and flying-electron qubits. Even within these established modalities, however, many required components and enabling technologies continue to be developed in a modality-specific and fragmented manner.

Looking ahead, component-development platforms are likely to become more consolidated, enabling more efficient R\&D and manufacturing scale-up. In parallel, continued progress in standardization should allow a broader set of component manufacturers to develop and supply parts across multiple modalities. For example, while cryogenic systems and interconnect technologies are already being advanced through largely shared approaches across several modalities, similar cross-modality convergence is anticipated for other key subsystems, including chip carriers, lasers, photonic integrated circuits (PICs) based on silicon/thin-film lithium niobate (TFLN) photonics, and detectors (e.g., transition-edge sensors (TES) as photon-number-resolving detectors, superconducting nanowire single-photon detectors (SNSPDs), single-photon avalanche diodes (SPADs), and CMOS-based devices). This direction is essential not only to accelerate the adoption of improved technologies within each modality, but also to reduce cost and establish robust supply chains through broader industrial participation.

Modality-specific technologies are, of course, expected to be developed more efficiently---and to higher performance---on top of such shared development platforms. However, it remains desirable to identify commonalities across modalities as broadly as possible, so that components with shared requirements can be developed horizontally on cross-disciplinary, modality-agnostic platforms.

Examples that are comparatively less transferable across modalities include Josephson-junction technologies for superconducting-qubit quantum computing; on-chip micromagnet technologies for silicon spin quantum computing; and, for optical/photonic quantum computing, high-purity single-photon sources, quantum-interference-optimized elements within low-loss PICs, and high-performance, high-squeezing-level optical squeezers. For neutral-atom quantum computing, modality-specific items include ultra-high-/extreme-high-vacuum (UHV/XHV) glass-cell technologies, ultra-high-power and ultra-stable lasers, and large-scale optical-tweezer systems with high-speed, high-bandwidth beam control. For trapped-ion quantum computing, stringent vacuum packages (typically UHV; in some implementations approaching XHV) and the associated system-integration technologies for installing and operating multiple in-vacuum components are likewise central. It is essential that such modality-specific technologies be developed and matured within their respective modalities.

In this way, the quantum industry---encompassing both quantum computing and quantum sensing---is expected to continue to mature, enabling these technologies to deliver broad societal impact, underpinned by metrological precision measurements and international standardization.

\begin{acknowledgments}
The author gratefully acknowledges Yuichi Nakamura and Kazutomo Hasegawa for providing input on the organizational structure of IEC/ISO JTC~3. The discussion of NMI-Q pre-standardization activities, as well as related standardization efforts, benefited greatly from exchanges with Kevin Thomson (NRC, Canada), F\'elicien Schopfer (LNE, France), Barbara Goldstein (NIST, USA), Tim Prior (NPL, UK), Nicolas Spethmann (PTB, Germany), Jan Herrmann (NMIA, Australia), and Ivo Degiovanni (INRiM, Italy), with whom the author serves on the NMI-Q Steering Committee (SC). The author also gratefully acknowledges Chiharu Urano (AIST G-QuAT) for sharing technical insights on non-magnetic materials, Yasutaka Amagai (AIST G-QuAT) for providing input on diamond nitrogen-vacancy (NV) centers, and Hidekazu Hashimoto (NITE) for his valuable advice on conformity assessment schemes. The author sincerely appreciates their cooperation and support.

\end{acknowledgments}

\bibliographystyle{apsrev4-2}
\bibliography{Metrology_for_Quantum_rev}

\section*{Author biography}
Nobu-Hisa Kaneko earned his Ph.D. in condensed matter physics from Tohoku University in Sendai, Japan, in 1997. Between 1996 and 1999, he conducted research at the National Research Institute for Inorganic Materials in Tsukuba, Japan. In 1999, he joined the Department of Applied Physics at Stanford University as a postdoctoral researcher, later becoming a physicist at the Stanford Linear Accelerator Center (SLAC), Stanford University, in California, USA.\\
In 2003, he joined the National Metrology Institute of Japan (NMIJ), part of the National Institute of Advanced Industrial Science and Technology (AIST) in Tsukuba. His research at NMIJ/AIST has focused on precision measurements and quantum electrical standards, particularly through the quantum Hall effect, the Josephson effect, and single-electron tunneling effect. Over the years, he has held several leadership roles within the institute, including Section Chiefs, Division Head, and Group Leaders in the field of electricity and magnetism. Since 2017, he has served as a Principal Researcher, mentoring early-career scientists and guiding research strategy. His primary research interests lie in condensed matter and materials physics, especially their applications in metrology.\\
He also serves as a Principal Researcher and a Team Leader for G-QuAT  (Global Research and Development Center for Business by Quantum-AI Technology), a quantum research initiative launched by AIST to advance collaborative efforts in quantum-AI computing and quantum sensors. The program fosters partnerships with industry and public institutions internationally.\\
He is a Steering Committee Member of NMI-Q and a member of IEEE, JSAP, and IEEJ, and one of the Quantum 100 for UNESCO's International Year of Quantum Science and Technology.

\end{document}